# Intrinsic inhomogeneities and effects of resistive switching in doped manganites.

N.A.Tulina, L.S.Uspenskaya

*Institute of Solid State Physics RASc, Chernogolovka, Moscow dist., 142432, Russia*

V.V.Sirotkin

*Institute of Microelectronics Technology and High Purity Materials RASc, Chernogolovka, Moscow dist., 142432, Russia*

Y. M. Mukovskii, D.A Shulyatev

*Moscow State Steel and Alloys Institute, Moscow, Leninskii pr-t, 4, 119991,Russia*

The effect of resistive switching in doped manganites being in the ferromagnetic state has been studied using resistive and magneto-optic methods. The visualization of magnetic structure of $La_{0.75}Sr_{0.25}MnO_{3-x}$ single crystals, and its transformation under electric current proposed local superheating of the material above the Curie temperature, which was supported by numerical calculation. The obtained results suggest a significant role of micrometer-scale inhomogeneity of manganites in phase separation, magnetic and transport properties of the material.

Perovskite manganites structures with colossal magnetoresistance effect (CMR), exhibiting phase separation, are prospective for being employed in telecommunication as switches and logic elements and etc. Especially useful for applications is a possibility to control the properties of magnetic layer by external electric or magnetic field. It has been theoretically proposed and experimentally shown [1-7], that phase separation (PS) plays an essential role in physics of mixed-valent manganites. Two mechanisms were proposed to explain PS: electronic phase separation for nanometer-scale inhomogeneity, and the effects of disorder for micrometer-scale inhomogeneity [6,7]. Structure disorder, crystals imperfections, elastic strains and quenched disorder can induce PS on submicron scale [8-10].

The current-voltage characteristics (IVC) of doped manganites correspond to the phase composition of the material. In the region of ferromagnetic (FM) phase the IVC are linear (metallic state) whereas in the paramagnetic (PM) region and in the phase separation state they demonstrate a power dependence, reflecting an activation transport behavior [11,12]. Therefore, the FM to PM transition initiates the resistance switching from low to high resistance. This effect is reversible and nonpolar, as distinct from the polar current switchings with memory effects, observed on the normal metal–manganites interface [13,14].

The nature of the switching effects, their association with spin-dependent transport, electron drag or just with local superheating is under investigation. In this study, we applied magneto-optic visualization technique (MO) to reveal which of the above-cited scenarios yields the switching effect in manganites. Afterwards, we confirmed our conclusion by numerical simulation of the process.

The experiments were performed on $La_{0.75}Sr_{0.25}MnO_3$ single crystals grown by the floating zone technique [15]. As-grown samples were cylinders 12 mm long, 3-5 mm in diameter, with [100] crystallographic axis inclined about 10 degrees to the growth direction, sub-block mosaicity was about 0.5 degree. Local X-ray analysis has shown, that concentration of La, Sr, Mn in the samples corresponded to the nominal with accuracy of 6%. The samples were cut along the growth-axis, polished mechanically and etched electrochemically to prepare the flat polished surface, required for MO observations. Then dump-bells shaped of 1-3 mm in diameter and 5-7 mm in length were prepared. The bridges were used to determine the IVC and temperature dependence of the resistance, R(T), by four-probe method and to study magnetic structure transformation under the current flow. MO observations were performed in the temperature range of 34-300 K, under magnetic field up to 2000 Oe, and under electrical current up to 2A.

The transport and magnetic properties of as-grown samples were modified by annealing via heating of the samples with electrical current (until the sample begin to glow). We have found that such annealing modified cardinally magnetic structure of the material and its transport properties [16].

The value of resistance was increased by a few orders; temperature dependence of resistance R(T) was changed from metallic to semiconductive type; transition temperature from paramagnetic to ferromagnetic state, $T_c$, was decreased from 340 K down to 170 K, figure 1. The resistive switching under electrical current, i.e. the transition to non-linearity on IVC with current growth, was observed for all samples while they were in ferromagnetic state, but the hysteretic behavior in resistive switching became more apparent on the IVC of annealed samples, figure 2.

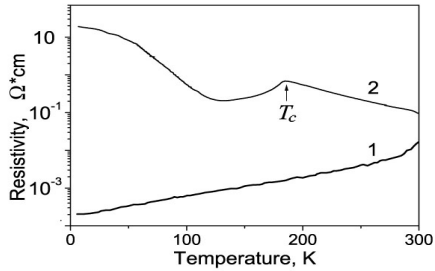

*Fig 1. Temperature dependence of resistivity of $La_{0.75}Sr_{0.25}MnO_{3-\delta}$ single crystals, curve 1- as-grown crystal, curve 2 - thermally treated crystal*

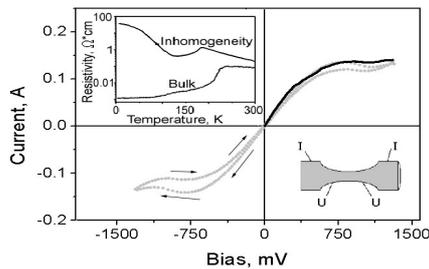

*Fig 2. Current-voltage characteristics of annealed $La_{0.75}Sr_{0.25}MnO_3$ measured at T = 78 K (open squares) and the numerical simulation of IVC (crosses); inset 1 – $\rho(T)$ for inhomogeneity and for the bulk, taken in numerical simulation; inset 2 - sketch of four-probe measurements on a*

The domain structure of the manganites was hardly recognized by MO before annealing, figure 3a. After fast annealing by electrical current, periodic magnetic domain structure appears with visible magnetic moment modulation in two directions, along and across pattern, figure 3b. Typical scale for this modulation was found to be ~1-50 μm. So, we have found that thermal heating stimulates micrometer scale PS, which was discussed in [7].

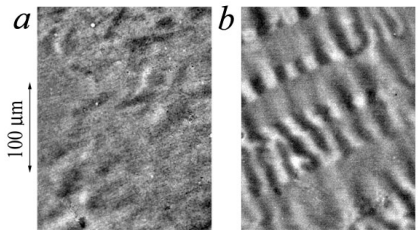

*Fig 3. Magnetic domain structure of the sample before and after heat treatment: diffused structure before annealing (a) is changed by regular domain structure (b)*

Direct magneto-optic observation of the effect of flowing current on transformation of the domain structure of manganites has shown that principal mechanism of current influence is local overheating of crystals, which causes subsequent local ferromagnetic-paramagnetic transition. This is illustrated by figures 4a-4d, where the transformation of ferromagnetic domain structure occurs under the growing current flow at T = 34 K. The ferromagnetic domains with perpendicular component of magnetization could be recognized on the pictures as stripes with black or white contrast. Areas with smaller magnetization look gray. Under the current flow the black-white contrast on the ferromagnetic domains is reduced into the dark-light gray contrast. This could be treated as transition from ferromagnetic to paramagnetic state due to direct current influence on magnetization state or due to heating caused by current. We did visualization of weak current paths in studied manganites, like it was done by M. Tokunaga et al [17] on LaPrCaMnO single crystals, subtracting images taken at opposite current, and have found, that remarkable current distortion takes place on the boundaries between gray and black/white domains.

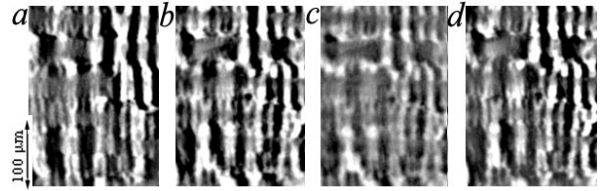

*Fig 4. Evolution of the domain structure of annealed $La_{0.75}Sr_{0.25}MnO_{3-x}$ at T = 34 K under the flowing current, I = 0, 350, 1150, 0 mA*

The local heating begins namely on these boundaries and spreads into ferromagnetic domains, figures 4a-4c. As a result, ferromagnetic domains are reducing in size, the spontaneous magnetization in the domains is also reducing and becomes inhomogeneous, with higher magnetization in the domain centers and lower magnetization closer to domain boundaries. So, more and more sample volume under the current becomes paramagnetic, compare figures 4a and 4c. After the current switching off domain structure restores, but not exactly the same, compare figures 4a and 4d.

Picture of sample overheating is even better seen in the series of subsequent images taken at higher temperature, at T = 80 K, while the current of 145 mA application, which is shown in series 1-2 in figure 5. Spontaneous domain structure seen in figure 5 (image 1, series1) was saturated by application of the longitudinal magnetic field of 0.2 T. After that, MO visualizes not domain structure, but spatial modulation of magnetization, which looks as bright and dark stripes, figure 5 (image 2, series1). This modulation corresponds either to the twin structure of manganites [18], or to some fine growth inhomogeneity, which is not recognized by X-ray microanalyses. The current flow first reduces magnetization, which is seen as a reduction of image contrast, figure 5 (image 3, series1) and finally turns out ferromagnetic phase into paramagnetic one. The boundary between ferromagnetic and pure paramagnetic phases is seen as bright stripe (marked by arrows in figure 5 (images 4-6, series1)) because of large magnetic field distortion. The boundary moves along the sample wiping away the magnetization modulation. Finally, the entire sample becomes paramagnetic, figure 5 (image 7, series1). Switching off of the current causes reverse process, which includes appearance of ferromagnetic state by inverse motion of the boundary and appearance of magnetic structure, figure 5 (series2). This restored structure practically coincides with the initial one, indicating that the observed magnetization modulation is caused by the fine inhomogeneity of crystal structure, which is still stable at $T \sim T_c$. The effect does not depend upon the current direction, but depend upon the current strength. The higher the current the faster this process, the faster the transition from ferromagnetic state to paramagnetic one. Without longitudinal magnetic field the same transition from ferromagnetic to paramagnetic state under the flowing current occurs. But restored after switching off of the current domain structure differ from the initial structure. Only longitudinal modulation of magnetization value practically coincides with initial one. The current, which is weaker than some threshold value, causes only local reversible transformation of magnetic structure, as was shown on figure 4.

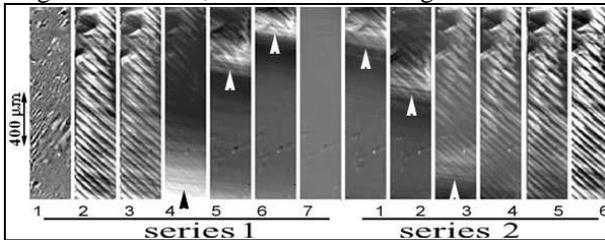

*Fig 5. Transformation of magnetic structure under current flow at T = 80 K: series 1 – (1) - spontaneous domain structure, (2) - ferromagnetic inhomogeneities visualized under $\mu_0 H = 0.2\ T$ and ferromagnetic to paramagnetic structure transformation (3 – 7) under I = 145 mA at t = 2, 5, 7, 9, 25 s after current turn on, series 2 – ferromagnetic structure restore from paramagnetic state at t = 12, 14, 21, 24, 28, 40 s after current turn off.*

The measured experimentally IVC were approximated numerically under the assumption of the inhomogeneous resistive properties of the crystal. Following the experimental observations we suggested an inhomogeneity with dimensions ($r_{i1}, r_{i2}, l_i$) is located in a bridge. This inhomogeneity was supposed to result from thermal treatment and from phase separation. Therefore it's resistive properties, $\rho_s(T)$, were modeled as the properties of manganite with lessened doping with respect to the bulk $\rho_b(T)$, shown on the inset in figures 2.

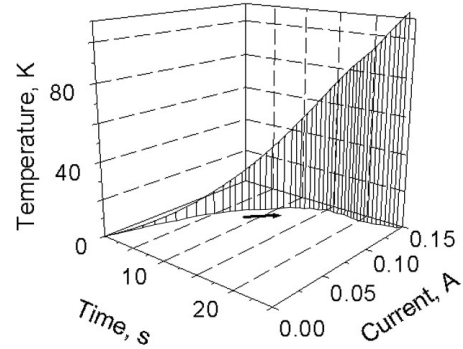

*Fig 6. Dependence of maximal overheating temperature upon the current strength and the time of it's flow*

We calculated the IVC, temperature fields, equipotential, and current distributions in the wire of radius $r_0$ with length L at different velocities of electric field sweep in the 2D approximation [19,20] based on equations:

$$\nabla[\sigma(P,T)\ \nabla\varphi]=0, \qquad (1)$$

$$C_p(P,T)\partial T/\partial t=\nabla[\lambda(P,T)\nabla T]+\sigma[\nabla\varphi]^2, \qquad (2)$$

where $\sigma$ is electric conductivity, $\varphi$ - electric potential, p=(r, z) - coordinate, $C_p$ - heat capacity, and $\lambda$ - heat conduction. The maximal voltage of the sweep V~1V, sweep time t ~25 sec. The temperature at the ends of the wire was assumed preset. Other parameters were T=78 K, $r_0$=4.2 *$10^{-2}$cm, L=5*$10^{-1}$cm, $r_{i1}$=4.2*$10^{-2}$cm, $r_{i2}$=0.5*$10^{-2}$cm, $l_i$=1.5*$10^{-1}$cm. Figures 2 and 6 demonstrate obtained approximation of the experimental data. The model with the embedded inhomogeneity is seen to describe fairly well the experiment. So, the calculation has confirmed the magneto-optic observations. The assumption about inhomogeneous state of manganites, i.e. about space variation of their resistive and magnetic properties, could be resulted in an inhomogeneous distribution of electric fields and current, and, consequently, in local superheating following by ferromagnetic to paramagnetic transition and switching effects. The switching could be reversible and would not lead to the total collapse of ferromagnetic structure, if the regions with overheating are small enough, which is confirmed by small hysteresis of the IVC. Taking into account, the chemical composition of the annealed material in La, Sr, Mn elements is unchanged, as was shown by local X-ray analysis, the origin of observed by MO micrometer-scale inhomogeneities in annealed by electrical current manganites bridges could be understand in terms of anisotropic strains and induced by the strains martensitic transformation [8,9].

Summarizing, it is shown by visualization of magnetic domain structure, visualization of weak current paths, observation of local material overheating and numerical simulation of the processes, that the effect of the IVC switching in the ferromagnetic phase region is rather associated with superheating of a local sample fractions up to Curie temperature, than by electron instability effect in doped manganites. It is shown, that even small inhomogeneity on micro-meter scale with the resistivity different from the bulk one, causes local superheating of the bulk part, embedded into locally superheated fine parts and gives rise to the current-voltage characteristic switching over from the linear current-voltage dependence to the nonlinear one and to related hysteresis phenomena.

This work has been supported by RFBR grants 05-02-17175a and by Programs of RASc "Properties of Condensed Matter" and "New materials and structures".

1. Dagotto E. , Reviews of Modern Physics. **66** (1994) 763;
2. Нагаев Е.Л. , УФН. **166** (1996) 833;
3. Горьков Л.П.. УФН. **168** (1998) 665;
4. Coey M.D., Viret M., S. von Molnar. // Advances in Physics. **48** (1999) 167;
5. Dagotto E., Hotta T., Moreo A. , Physics Reports. **344** (2001) 1;
6. Salamon M.B. , Reviews of Modern Physics. **73** (2001) 583.
7. Нагаев Е.Л. Physics Reports **346** (2001) 531.
8. Uera M. and S.-W.Cheong, Europhysics Lett **52** (2000) 674.
9. Podzorov.V.,B.G.Kim et al., Phys. Review B **64** (2001) 140406(R)
10. Ahn K.N.,T.Lookman,A.R.Bishop,Letter to Nature **428** (2004) 401.
11. Tulina N.A., Zver'kov S.A., Mukovskii Y.M., et al. Europhysics Lett **56** (2001) 836.
12. Yuzhelevski Y. et al., Phys. Review B **64** (2001) 224428.
13. Tulina N.A., Zver'kov S.A., Arsenov A, et al. Physica C **385** (2003) 563;
14. Fors R., S. I. Khartsev, and A. M. Grishin , Phys. Review B **71** (2005) 045305
15. Shulyatev D., Karabashev S., Arsenov A., Mukovskii Ya., J. Crystal Growth. **198/199** (1999) 511.
16. Tulina N.A., Uspenskaya L.S., Shulyatev D.A., Mukovskii Ya.M. to be published in Russian Journal "Izvestiya RAN", 2006
17. M.Tokunaga, Y.Tokunaga, and T.Tamegaki, Phys. Review Let**. 93** ( 2004) 037203.
18. Khapikov A., Uspenskaya L., Bdikin I, et al., Appl. Phys. Lett. **77** (2000) 2376.
19. Tulina N.A., V.V.Sirotkin, Physica C **400** (2004) 105.
20. Sirotkin V. V., Computers Math. Applic. **38** no. 3/4 (1999) 119.